\documentclass[12pt]{article}

\usepackage{array, booktabs}
\usepackage{amsmath,amssymb,amsfonts,graphics,graphicx,amscd,amsfonts,epsfig,color}
\usepackage{subfig}
\usepackage{here}
\usepackage{cite}

\newcommand {\beq}{\begin{equation}}
\newcommand {\eeq}{\end{equation}}
\newcommand {\beqa}{\begin{eqnarray}}
\newcommand {\eeqa}{\end{eqnarray}}

\renewcommand{\theequation}{\thesection.\arabic{equation}}

\begin{document}
\setlength{\oddsidemargin}{0cm}
\setlength{\baselineskip}{7mm}

\begin{titlepage}
\renewcommand{\thefootnote}{\fnsymbol{footnote}}
\begin{normalsize}
\begin{flushright}
\begin{tabular}{l}
March 2017
\end{tabular}
\end{flushright}
\end{normalsize}

~~\\

\vspace*{0cm}
\begin{Large}
\begin{center}
{Correlation functions and renormalization in a scalar field theory \\
on the fuzzy sphere}
\end{center}
\end{Large}
\vspace{1cm}

\begin{center}
Kohta H{\sc atakeyama}\footnote
{
e-mail address : 
hatakeyama.kohta.15@shizuoka.ac.jp}
{\sc and}
Asato T{\sc suchiya}\footnote
{
e-mail address : tsuchiya.asato@shizuoka.ac.jp}\\
\vspace{1cm}

{\it Department of Physics, Shizuoka University}\\
{\it 836 Ohya, Suruga-ku, Shizuoka 422-8529, Japan}\\
\vspace{0.3cm} 
\end{center}

\hspace{5cm}

\begin{abstract}
\noindent
We study renormalization in a scalar field theory on the fuzzy sphere. The theory is realized by a matrix model, where
the matrix size plays the role of a UV cutoff.
We define correlation functions by using 
the Berezin symbol identified with a field and calculate them nonperturbatively
by Monte Carlo simulation.
We find that the 2-point and 4-point functions 
are made independent of the matrix size
by tuning a parameter and performing a wave function renormalization.
The results strongly suggest that the theory is 
nonperturbatively renormalizable in the ordinary sense.
\end{abstract}
\vfill
\end{titlepage}
\vfil\eject

\setcounter{footnote}{0}

\section{Introduction}
\setcounter{equation}{0}
\renewcommand{\thefootnote}{\arabic{footnote}} 

It is conjectured that noncommutative geometry plays an essential role
in the quantum theory of gravity. Indeed, it appears in various contexts of string 
theory (for a review, see \cite{Douglas:2001ba}.). 
For instance, field theories in noncommutative spaces are realized
\cite{Connes:1997cr,Aoki:1999vr} in
the matrix models \cite{BFSS, IKKT, DVV}, which are proposals for
nonperturbative formulation of 
string theory.
Thus it is important to elucidate how field theories in noncommutative spaces differ from those in ordinary spaces.
For this purpose, one needs to identify the behavior of basic quantities in
field theories such as correlation functions.

One of the most important features of field theories in noncommutative 
spaces is that the product for fields is noncommutative and nonlocal.
It yields IR divergences in perturbative expansion that originate from UV divergences.
This phenomenon is called UV/IR mixing \cite{Minwalla:1999px}and is known to be an obstacle to perturbative renormalization.

In this paper, we study multi-point correlation functions
in a typical and simple example of a field theory in noncommutative
spaces, a scalar field theory on the fuzzy sphere \cite{Madore:1991bw}.
The theory is given by a matrix model, 
where the matrix size $N$
plays the role of a UV cutoff.
There are several scaling limits where $N\rightarrow\infty$, corresponding
to the continuum limits.
Here, as a first approach to the above issue, it is reasonable to
consider the so-called commutative limit where $N\rightarrow\infty$
with the radius of the sphere fixed, since the theory obtained in this limit is 
expected to be closest to the theory on the ordinary sphere.
Indeed, as reviewed later, the former reduces to the latter in this limit at the 
classical (tree) level.
However, it was shown in \cite{Chu:2001xi,Steinacker:2016nsc}
that the one-loop contribution to the self-energy in this limit is not 
IR divergent but differs by a finite and nonlocal term
from that in the theory on the ordinary sphere.
This difference is called the UV/IR anomaly and is a finite analog of
the UV/IR mixing.
It is important to see whether this sort of difference exists nonperturbatively
or not, because it is possible that nonperturbative aspects of noncommutative 
field theory are relevant for quantum gravity.

Thus we calculate the correlation functions nonperturbatively by a
performing Monte Carlo simulation (for a Monte Carlo study of the model,
see \cite{Martin:2004un,Panero:2006bx,Panero:2006cs,
GarciaFlores:2009hf,Das:2007gm}. For a related analytic study of the model,
see \cite{Kawamoto:2015qla,Vaidya:2003ew,OConnor:2007ibg,Nair:2011ux,Polychronakos:2013nca,
Tekel:2013vz,Saemann:2014pca,Tekel:2014bta,Tekel:2015zga}.).
Here, in particular, we focus on renormalization, one of the most basic properties 
of field theories. We will see whether the theory is renormalized in the ordinary
manner; namely, whether 
the multi-point correlation functions become independent of the UV cutoff
$N$ if some parameters are tuned and a wave function renormalization 
is performed. Such nonperturbative renormalization in a nonlocal field theory 
should be nontrivial\footnote{Proof of perturbative renormalizability still seems to be
missing, while the theory in the commutative limit is naively considered to be
perturbatively renormalizable 
because the one-loop self-energy is the only diagram that is UV divergent
in the corresponding scalar field theory on the ordinary sphere.}.
The authors of \cite{Bietenholz:2004xs,Mejia-Diaz:2014lza} examined
the dispersion relation and so on in scalar field theories 
on the noncommutative torus by
calculating the 2-point correlation functions nonperturbatively
by Monte Carlo simulation
and concluded that
the theories are nonperturbatively renormalizable 
in the double scaling limit 
where the continuum and thermodynamic limits are
simultaneously taken at fixed noncommutative tensor.
The theories obtained in the double scaling limit 
are obviously different from field theories in ordinary spaces.
A similar analysis for gauge theories on 
the noncommutative torus was performed 
in \cite{Bietenholz:2002ch,Bietenholz:2006cz}.

Here, we find that the 2-point and 4-point correlation functions are
independent of $N$ if a parameter is tuned and a
wave function renormalization is performed.
These results strongly suggest that the theory is 
nonperturbatively renormalizable 
in the ordinary sense
and enable us to conjecture that 
the theory is specified by a parameter.
To support this conjecture, we examine the theory at a fixed $N$.
This is the first Monte Carlo study of renormalization on the fuzzy sphere.

To define the correlation functions, we regard 
the Berezin symbol \cite{Berezin:1974du} of the matrix constructed 
from the Bloch coherent state \cite{Gazeau} as a field.
As far as we know, the coherent state is used for the first time
in Monte Carlo study of noncommutative field theories\footnote{
For perturbative calculation
using 
the coherent state, see \cite{Iso:2000ew,Steinacker:2016nsc}.
Also note that
in \cite{Okuno:2015kuc,Suzuki:2016sca}
the coherent state is implicitly used for the calculation
of entanglement entropy on the fuzzy sphere by Monte Carlo simulation,
following the observation in \cite{Karczmarek:2013jca,Sabella-Garnier:2014fda}.}.
Thus another aim of this paper is to demonstrate that
the method developed here is a powerful one 
for Monte Carlo study of noncommutative
field theories.
We expect the method to be applied not only to the other limits of the theory but
also to other field theories in noncommutative spaces.

This paper is organized as follows.
In section 2, we review a scalar field theory on the fuzzy sphere,
which is realized by a matrix model. We compare it with a corresponding
field theory on the sphere by identifying the Berezin symbol constructed
from the matrix with the field.
In section 3, we define the correlation functions that we calculate by 
Monte Carlo simulation, and show the results of the simulations.
Section 4 is devoted to the conclusion and discussion.
In appendix A, we review the Bloch coherent state, the Berezin symbol and
the star product on the fuzzy sphere.
In appendix B, we show the results for the 1-point functions.

\section{Scalar field theory on the fuzzy sphere}
\setcounter{equation}{0}
Let us consider a scalar field theory on a sphere with the radius $R$:
\begin{align}
S_C=\frac{R^2}{4\pi}\int d\Omega \left(-\frac{1}{2R^2}({\cal L}_i\phi)^2
+\frac{\mu^2}{2}\phi^2+\frac{\lambda}{4}\phi^4\right) \ ,
\label{continuum action}
\end{align}
where
${\cal L}_i$ $(i=1,2,3)$ are the orbital angular momentum operators
and $d\Omega$ is the invariant measure on the sphere.
We parametrize the sphere by the standard polar coordinates $(\theta,\varphi)$.
Then ${\cal L}_i$ and $d\Omega$ take the following form:
\begin{align}
{\cal L}_{\pm} &\equiv {\cal L}_1 \pm i {\cal L}_2 
=e^{\pm i\varphi} \left( \pm \frac{\partial}{\partial \theta} 
+i \cot \theta \frac{\partial}{\partial \varphi} \right) \ ,\nonumber\\
{\cal L}_3 &= -i \frac{\partial}{\partial \varphi} \ ,
\end{align}
and $d\Omega=\sin\theta d\theta d\varphi$.

A noncommutative counterpart of (\ref{continuum action}) is 
given by a matrix model:
\begin{align}
S=\frac{R^2}{2j+1}\mbox{Tr}\left(-\frac{1}{2R^2}[L_i,\Phi]^2+\frac{\mu^2}{2}\Phi^2
+\frac{\lambda}{4}\Phi^4\right) \ ,
\label{action}
\end{align}
where $j$ is a non-negative integer or half-integer, and 
$\Phi$ is a $(2j+1)\times (2j+1)$ Hermitian matrix.
$L_i$ are the generators of the $SU(2)$ algebra with the spin $j$ representation,
obeying the commutation relation $[L_i,L_j]=i\epsilon_{ijk}L_k$.
$j$ plays the role of a UV cutoff.
We also denote the matrix size by $N$; namely, $N=2j+1$.

In this paper, we are concerned with the so-called commutative limit, where
$N\rightarrow\infty$ with $R$ fixed.
Hereafter, we put $R=1$ without loss of generality.
We briefly review below that
(\ref{action}) reduces to (\ref{continuum action}) at the classical level in this limit
while the former differs from the latter due to the UV/IR anomaly at the quantum
level.

Here, in order to see the correspondence between the above two theories,
we introduce the Bloch coherent state\cite{Gazeau}\footnote{See also
\cite{Alexanian:2000uz,Hammou:2001cc,Presnajder:1999ky,Ishiki:2015saa}} and
the Berezin symbol\cite{Berezin:1974du}. The Bloch coherent state denoted by 
$|\Omega \rangle$ ($\Omega=(\theta,\varphi)$)
is localized around the point $(\theta,\varphi)$ on the sphere.
The basic properties of the Bloch coherent state are reviewed in appendix A.

The Berezin symbol for an $N\times N$ matrix $A$ is defined by
\begin{align}
f_{A}(\Omega)=\langle\Omega|A|\Omega\rangle \ .
\end{align}
By using (\ref{explicit form}), one can easily show that 
\begin{align}
f_{[L_i,A] }(\Omega)={\cal L}_i f_{A}(\Omega) \ .
\label{derivative}
\end{align}
Also, (\ref{property4}) implies that
\begin{align}
\frac{1}{2j+1}\mbox{Tr}(A)=\int \frac{d\Omega}{4\pi} f_A(\Omega) \ .
\label{trace and integral}
\end{align}

The star product for the Berezin symbols is defined by
\begin{align}
f_A\star f_B (\Omega) = \langle \Omega | AB | \Omega \rangle \ ,
\label{Berezin symbol}
\end{align}
where $A$ and $B$ are $N\times N$ matrices.
To express the star product in terms of the Berezin symbol, we use
the stereographic projection given by
\begin{align}
z=\tan\frac{\theta}{2} e^{i\varphi} 
\end{align}
and denote the Bloch coherent state $|\Omega\rangle$ by $|z\rangle$
and the Berezin symbol $f_A(\Omega)$ by $f_A(z,\bar{z})$.
Then, the star product is expressed as
\begin{align}
f_A\star f_B(w,\bar{w})
=\frac{2j+1}{4\pi}4\int \frac{d^2z}{(1+|z|^2)^2}
(e^{-w\frac{\partial}{\partial z}}e^{z\frac{\partial}{\partial w}}
f_A(w,\bar{w}) )
(e^{-\bar{w}\frac{\partial}{\partial \bar{z}}}e^{\bar{z}\frac{\partial}{\partial \bar{w}}}
f_B(w,\bar{w}))
|\langle w | z \rangle |^2 
\label{expression for star product}
\end{align}
as shown in appendix A.
This shows that the star product is nonlocal and noncommutative.
Furthermore, it is easy to show that in the $j\rightarrow\infty$ limit
\begin{align}
\frac{2j+1}{4\pi}\frac{4}{(1+|z|^2)^2}
|\langle w | z \rangle |^2 
\rightarrow \delta^2(z-w) \ .
\end{align}
This implies that in the $j\rightarrow\infty$ limit the star product
reduces to the ordinary product. Namely,
\begin{align}
f_A\star f_B(w,\bar{w}) \rightarrow f_A(w,\bar{w})f_B(w,\bar{w}) 
\end{align}
or
\begin{align}
f_A\star f_B(\Omega) \rightarrow f_A(\Omega)f_B(\Omega) \ .
\label{ordinary product}
\end{align}

(\ref{derivative}), (\ref{trace and integral}) and (\ref{ordinary product}) show that
the theory (\ref{action}) reduces to the one (\ref{continuum action}) in the 
commutative ($N\rightarrow\infty$) limit at the classical (tree) level
if $f_{\Phi}(\Omega)$ is identified
with $\phi(\Omega)$.
However, as shown to the one-loop order 
in \cite{Chu:2001xi,Steinacker:2016nsc}, (\ref{action}) differs 
from (\ref{continuum action}) by a finite and nonlocal 
term because the UV cutoff $N$ must be kept finite in calculating the radiative corrections.
Namely, the quantization and the commutative limit are not commutative.
This phenomenon is called the UV/IR anomaly.

\section{Correlation functions}
\setcounter{equation}{0}
\subsection{Definition of correlation functions}
For later convenience,
we introduce a shorthand notation for the Berezin symbol:
\begin{align}
\varphi(\Omega) = f_{\Phi}(\Omega)=\langle\Omega | \Phi | \Omega\rangle \ .
\end{align}
In the theory (\ref{action}), the $n$-point correlation function is defined by
\begin{align}
\left\langle \varphi(\Omega_1)\varphi(\Omega_2)\cdots\varphi(\Omega_n)
\right\rangle
=\frac{\int d\Phi \ \varphi(\Omega_1)\varphi(\Omega_2)\cdots\varphi(\Omega_n) 
\ e^{-S}}{\int d\Phi \ e^{-S}} \ ,
\label{n-point function}
\end{align}
where
\begin{align}
d\Phi=\prod_{i=1}^N d\Phi_{ii} \prod_{1\leq j < k\leq N}
d\mbox{Re}\Phi_{jk} d\mbox{Im}\Phi_{jk} \ .
\end{align}
The correlation function (\ref{n-point function}) is an analog of
$\langle \phi(\Omega_1)\phi(\Omega_2)\cdots\phi(\Omega_n)\rangle$ in
the theory (\ref{continuum action}).

Suppose that the matrix $\Phi$ in (\ref{action}) is renormalized as
\begin{align}
\Phi=\sqrt{Z}\Phi_r \ ,
\end{align}
where $Z$ is a factor of the wave function renormalization, and $\Phi_r$ is the
renormalized matrix.
Correspondingly, the renormalized Berezin symbol $\varphi_r(\Omega)$ is defined by
\begin{align}
\varphi(\Omega)=\sqrt{Z}\varphi_r(\Omega) \ ,
\end{align}
and the renormalized $n$-point correlation function
$\left\langle \varphi_r(\Omega_1)\varphi_r(\Omega_2)\cdots\varphi_r(\Omega_n)
\right\rangle$ is defined by
\begin{align}
\left\langle \varphi(\Omega_1)\varphi(\Omega_2)\cdots\varphi(\Omega_n)
\right\rangle
=Z^{\frac{n}{2}}
\left\langle \varphi_r(\Omega_1)\varphi_r(\Omega_2)\cdots\varphi_r(\Omega_n)
\right\rangle \ .
\end{align}

\begin{figure}
\begin{center}
\includegraphics[width=5cm]{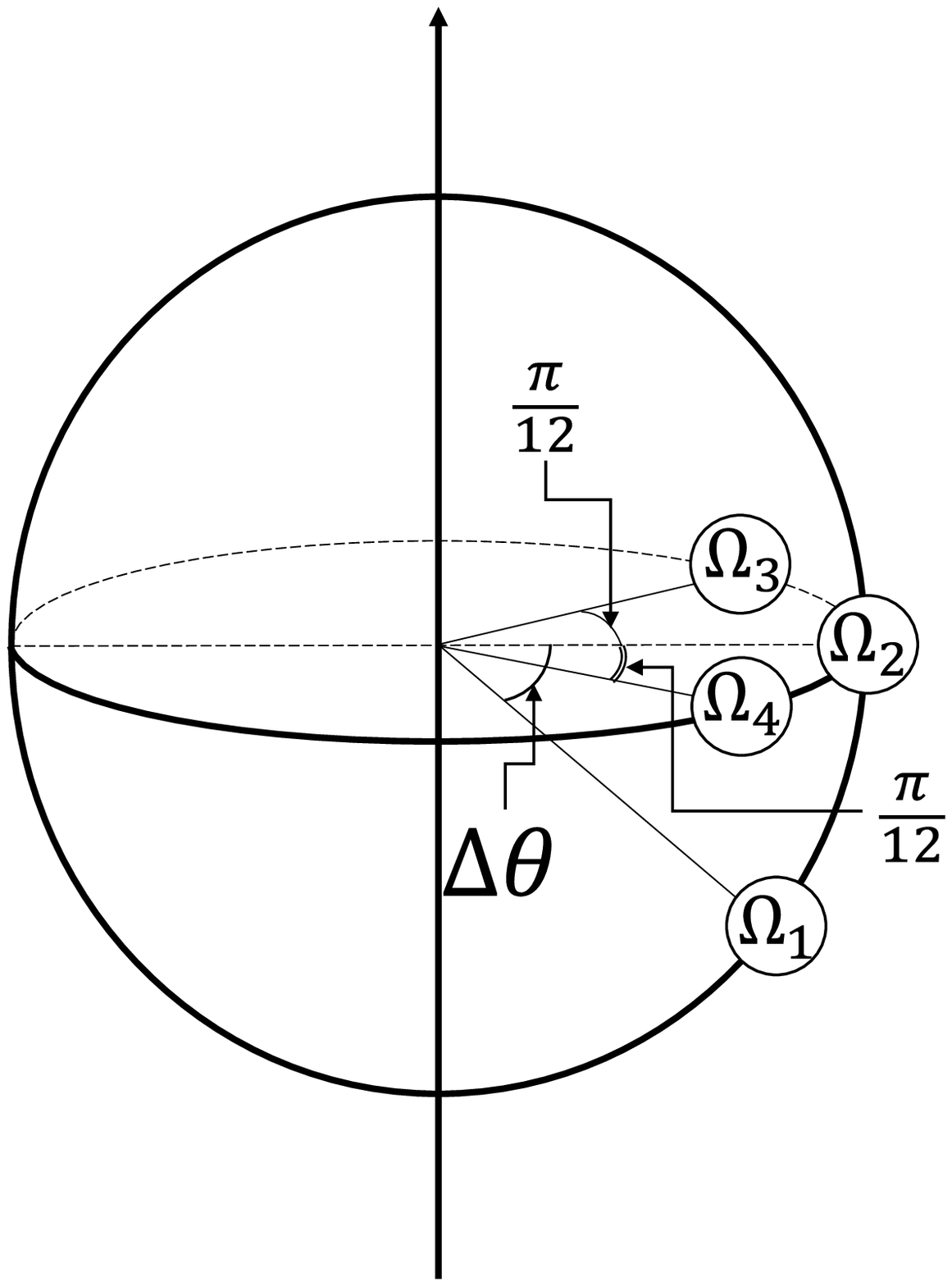}
\caption{Four points on the sphere selected for the correlation functions. }
\label{sphere}
\end{center}
\end{figure}

In the following, we calculate the following correlation functions by Monte Carlo
simulation:
\begin{align}
&\mbox{1-point function:} \; \left\langle \varphi(\Omega_1) \right\rangle \ ,
\nonumber\\
&\mbox{2-point function:} \; \left\langle \varphi(\Omega_p)\varphi(\Omega_q)
\right\rangle \; (1\leq p < q\leq 4) \ , \nonumber\\
&\mbox{4-point function:} \; \left\langle \varphi(\Omega_1)\varphi(\Omega_2)
\varphi(\Omega_3)\varphi(\Omega_4) \right\rangle \ .
\label{correlation functions}
\end{align}
We show in appendix B that the 1-point functions vanish.
Thus, the 2-point function is itself the connected one, while
the connected 4-point function is defined by
\begin{align}
\left\langle \varphi(\Omega_{1}) \varphi(\Omega_{2}) 
\varphi(\Omega_{3}) \varphi(\Omega_{4}) \right\rangle_c 
=&
\left\langle \varphi (\Omega_{1}) \varphi (\Omega_{2}) \varphi
(\Omega_{3}) \varphi (\Omega_{4}) \right\rangle 
- \left\langle \varphi(\Omega_{1}) \varphi(\Omega_{2}) \right\rangle 
\left\langle \varphi(\Omega_{3}) \varphi(\Omega_{4}) \right\rangle \nonumber \\
&- \left\langle \varphi(\Omega_{1}) \varphi(\Omega_{3}) \right\rangle 
\left\langle \varphi(\Omega_{2}) \varphi(\Omega_{4}) \right\rangle 
- \left\langle \varphi(\Omega_{1}) \varphi(\Omega_{4}) \right\rangle 
\left\langle \varphi(\Omega_{2}) \varphi(\Omega_{3}) \right\rangle \ .
\end{align}
The renormalized correlation functions are defined as
\begin{align}
\left\langle \varphi(\Omega_{1}) \right\rangle 
&= \sqrt{Z} \left\langle \varphi_r(\Omega_{1}) \right\rangle \ , \\
\left\langle \varphi(\Omega_{p}) \varphi(\Omega_{q}) \right\rangle 
&= Z \left\langle \varphi_r (\Omega_{p}) \varphi_r (\Omega_{q}) \right\rangle \ , \\
\left\langle \varphi(\Omega_{1}) \varphi(\Omega_{2}) 
\varphi(\Omega_{3}) \varphi(\Omega_{4}) \right\rangle_c 
&= Z^2 \left\langle \varphi_r (\Omega_{1}) \varphi_r (\Omega_{2}) 
\varphi_r (\Omega_{3}) \varphi_r (\Omega_{4}) \right\rangle_c \ .
\end{align}
We choose $\Omega_p=(\theta_p,\varphi_p)$ $(p=1,2,3,4)$ as follows 
(see Fig.\ref{sphere}):
\begin{align}
\Omega_1 &= \left(\frac{\pi}{2} +\Delta \theta, \ 0\right) \ , \nonumber\\
\Omega_2 &= \left(\frac{\pi}{2} , \ 0\right) \ , \nonumber\\
\Omega_3 &= \left(\frac{\pi}{2} , \ \frac{\pi}{12}\right) \ , \nonumber\\
\Omega_4 &= \left(\frac{\pi}{2} , \ -\frac{\pi}{12}\right) \ ,
\end{align}
where $\Delta\theta$ is taken from $0.3$ to $1.5$ in steps of $0.1$.

\subsection{Renormalization}
We use the hybrid Monte Carlo method to calculate the correlation functions
(\ref{correlation functions}).

\begin{figure}[t]
\begin{center}
\includegraphics[width=11.0cm]{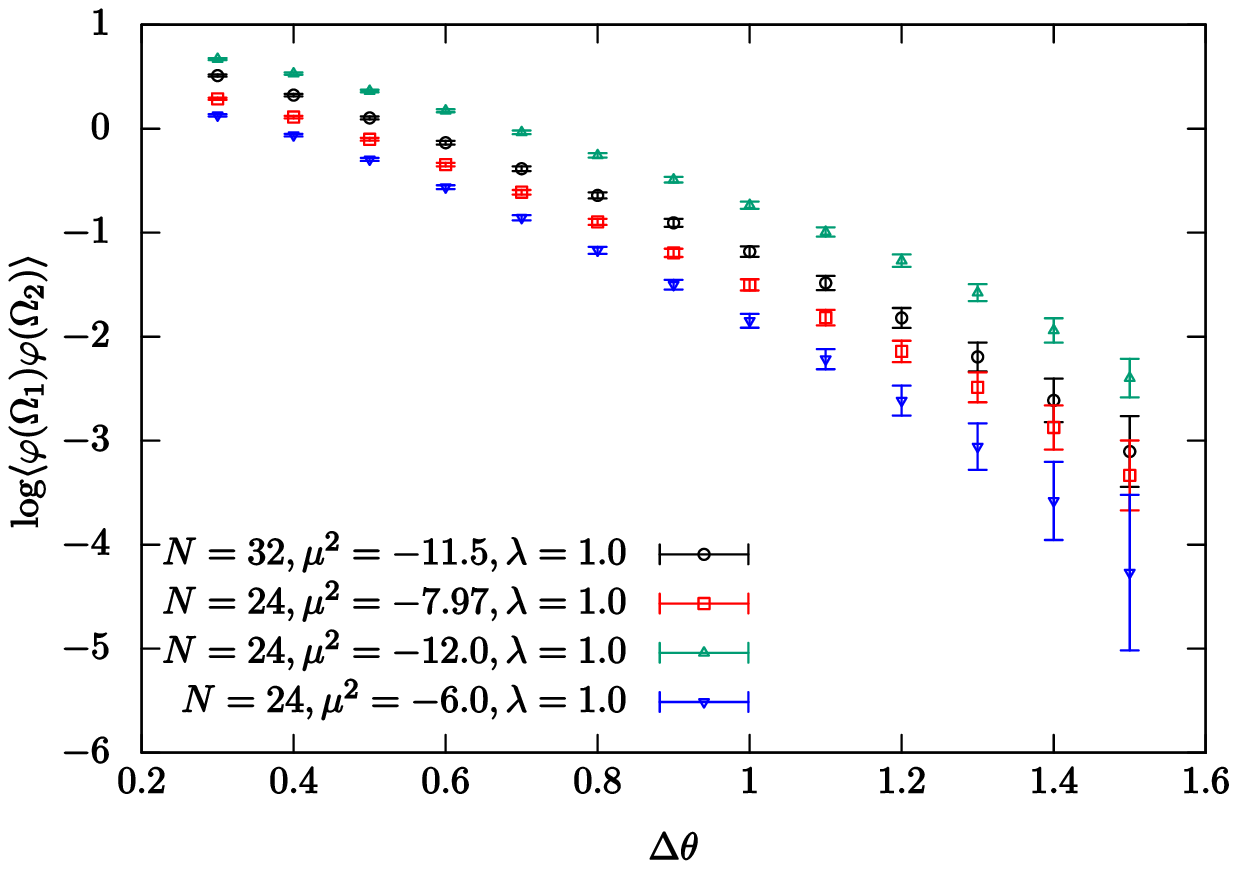}
\caption{$\log\left\langle\varphi(\Omega_1)\varphi(\Omega_2)\right\rangle$
at $\lambda=1.0$ is
plotted against $\Delta\theta$. The data for $N=32$ and $\mu^2=-11.5$ are 
represented by the circles, while the data for
$N=24$ and $\mu^2=-7.97, -12.0, -6.0$ are represented by
the squares, the triangles and the inverted triangles, respectively. }
\label{log_2pt_24}
\end{center}
\end{figure}

\begin{figure}[h]
\begin{center}
\includegraphics[width=11.0cm]{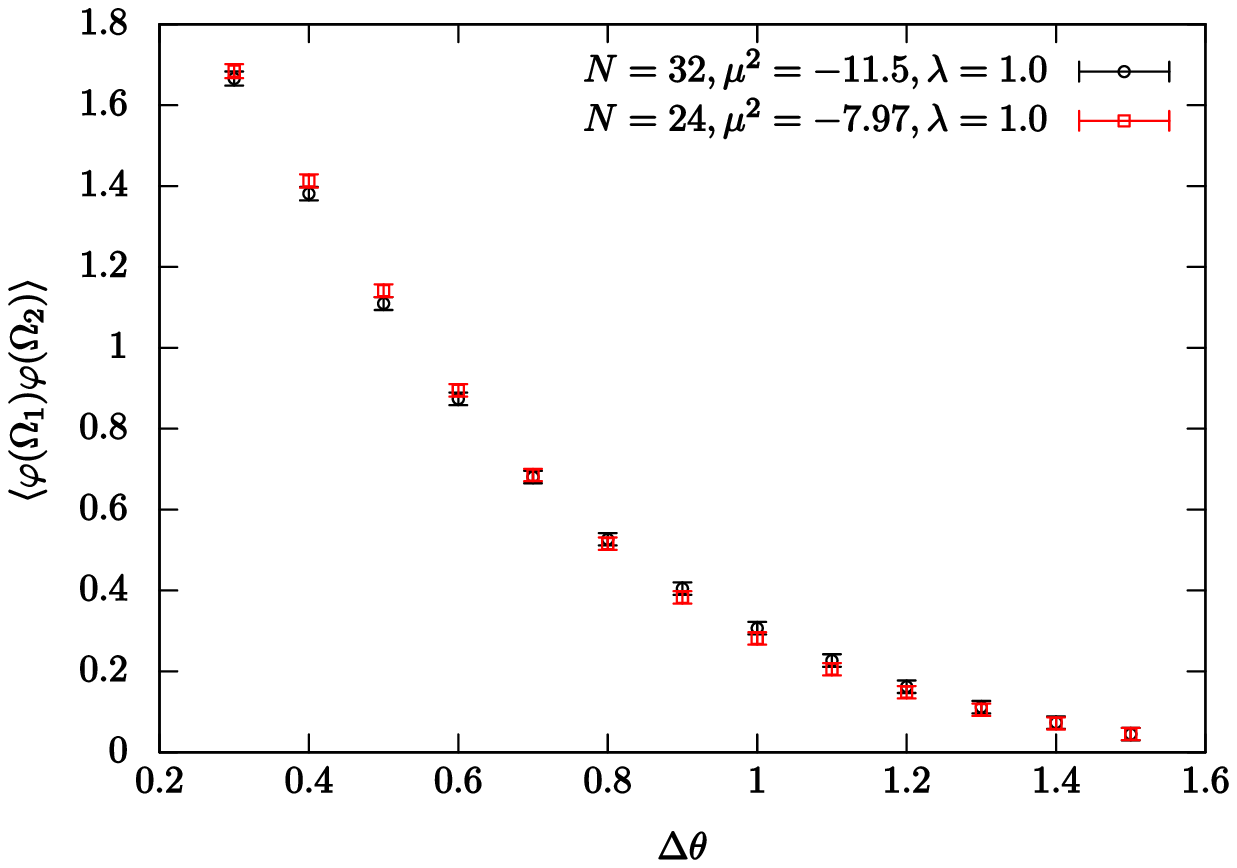}
\caption{$\left\langle\varphi(\Omega_1)\varphi(\Omega_2)\right\rangle$
at $N=32$, $\mu^2=-11.5$ and $\lambda=1.0$ is 
plotted against $\Delta\theta$, where the data are represented by the circles.
$\zeta_{24\rightarrow 32}
\left\langle\varphi(\Omega_1)\varphi(\Omega_2)\right\rangle$
with $\zeta_{24\rightarrow 32}=1.263$
at $N=24$, $\mu^2=-7.97$ and $\lambda=1.0$ 
is also plotted against $\Delta\theta$, where the data are represented by
the squares.}
\label{exp_2pt_24}
\end{center}
\end{figure}

First, we simulate at $N=32$, $\mu^2=-11.5$ and $\lambda=1.0$.
Then, keeping $\lambda=1.0$, we simulate at $N=24$ and
various values of $\mu^2$.
In Fig.\ref{log_2pt_24}, we plot
\begin{align} 
\log\left\langle\varphi(\Omega_1)\varphi(\Omega_2)\right\rangle
=\log Z +\log \left\langle \varphi_r(\Omega_1)\varphi_r(\Omega_2) \right\rangle
\end{align}
against $\Delta\theta$ at $N=32$ and $\mu^2=-11.5$
and at $N=24$ and typical values of $\mu^2$, $-7.97, -12.0, -6.0$.
We see that the data for $N=24$ and $\mu^2=-7.97$ agree with those
for $N=32$ and $\mu^2=-11.5$ if the former are simultaneously shifted 
in the vertical direction and that this is not the case for
the data for $N=24$ and $\mu^2=-12.0, -6.0$.
This implies that the renormalized 2-point function at $N=24$ and $\mu^2=-7.97$ 
agrees with that at $N=32$ and $\mu^2=-11.5$ and that we can determine
\begin{align}
\alpha_{24\rightarrow 32}\equiv
\log\left( \frac{Z(32)}{Z(24)} \right) \ .
\end{align}
Indeed, by using the least-squares method, we obtain
$\alpha_{24\rightarrow 32}=0.2334$ with the error $\delta\alpha_{24\rightarrow 32}
=0.0108$. In Fig.\ref{exp_2pt_24}, we plot 
$\left\langle\varphi(\Omega_1)\varphi(\Omega_2)\right\rangle$ 
at $N=32$ and $\mu^2=-11.5$ and 
$\zeta_{24\rightarrow 32}
\left\langle\varphi(\Omega_1)\varphi(\Omega_2)\right\rangle$ at
$N=24$ and $\mu^2=-7.97$ against $\Delta\theta$, where
\begin{align}
\zeta_{24\rightarrow 32}= e^{\alpha_{24\rightarrow 32}}=\frac{Z(32)}{Z(24)}=1.263 \ .
\end{align}
We indeed see a good agreement between the data for $N=24$ and thoes for
$N=32$.

\begin{figure}[h]
\begin{center}
\includegraphics[width=11.0cm]{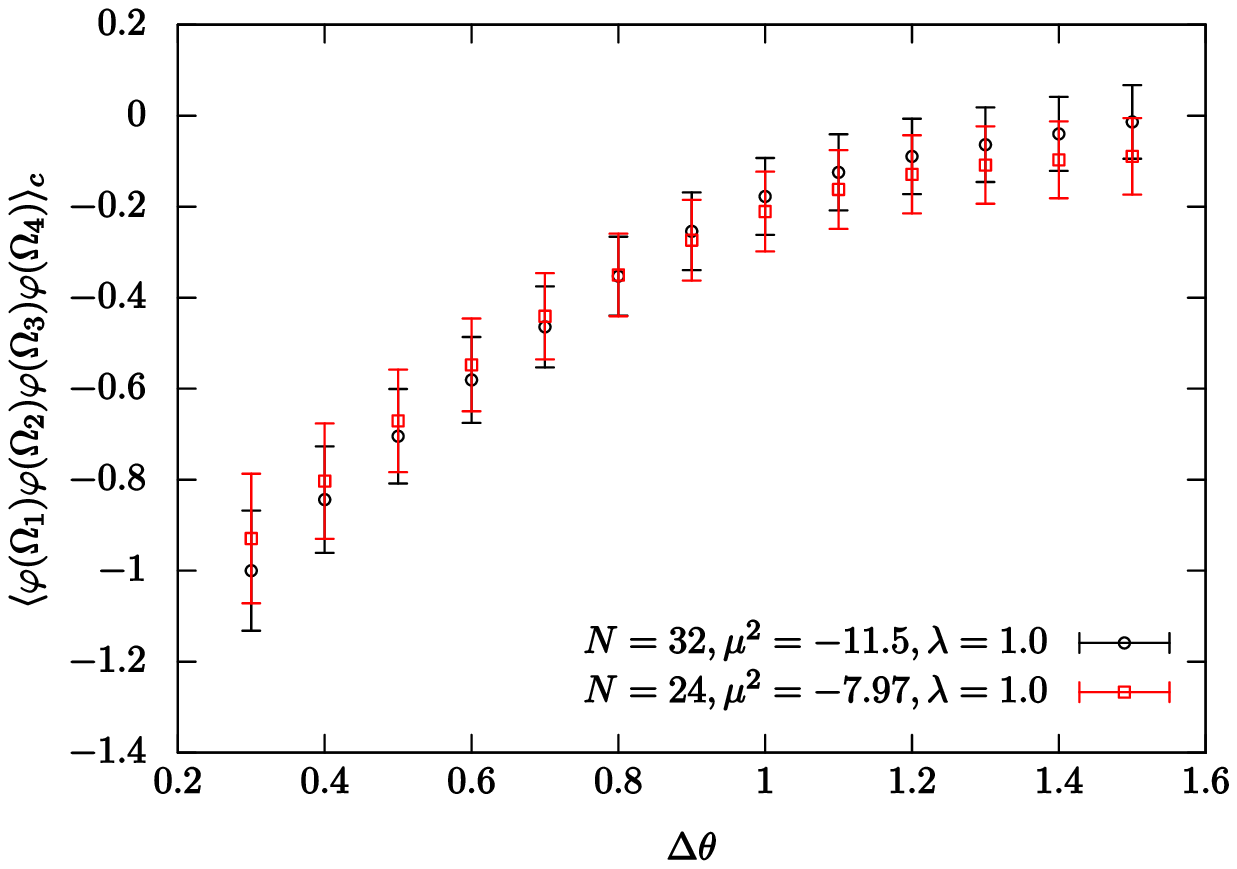}
\caption{$\left\langle \varphi(\Omega_{1}) \varphi(\Omega_{2}) 
\varphi(\Omega_{3}) \varphi(\Omega_{4}) \right\rangle_c$ 
at $N=32$, $\mu^2=-11.5$ and $\lambda=1.0$ is 
plotted against $\Delta\theta$, where the data are represented by the circles.
$\zeta_{24\rightarrow 32}^2
\left\langle \varphi(\Omega_{1}) \varphi(\Omega_{2}) 
\varphi(\Omega_{3}) \varphi(\Omega_{4}) \right\rangle_c$
with $\zeta_{24\rightarrow 32}^2=1.595$
at $N=24$, $\mu^2=-7.97$ and $\lambda=1.0$ 
is also plotted against $\Delta\theta$, where the data are represented by
the squares.}
\label{exp_4pt_24}
\end{center}
\end{figure}

Furthermore, we can expect that the renormalized connected 4-point function
at $N=24$ agrees with that at $N=32$.
Indeed, in Fig.\ref{exp_4pt_24}, we plot
$\left\langle \varphi(\Omega_{1}) \varphi(\Omega_{2}) 
\varphi(\Omega_{3}) \varphi(\Omega_{4}) \right\rangle_c$ at
$N=32$ and $\mu^2=-11.5$ and $\zeta_{24\rightarrow 32}^2
\left\langle \varphi(\Omega_{1}) \varphi(\Omega_{2}) 
\varphi(\Omega_{3}) \varphi(\Omega_{4}) \right\rangle_c$ 
at $N=24$ and $\mu^2=-7.97$ against $\Delta\theta$.
We again see a good agreement between the data for $N=24$ and thoes for
$N=32$, which implies that the renormalized connected 4-point function at $N=24$
agrees with that at $N=32$.

\begin{figure}[h]
\begin{center}
\includegraphics[width=11.0cm]{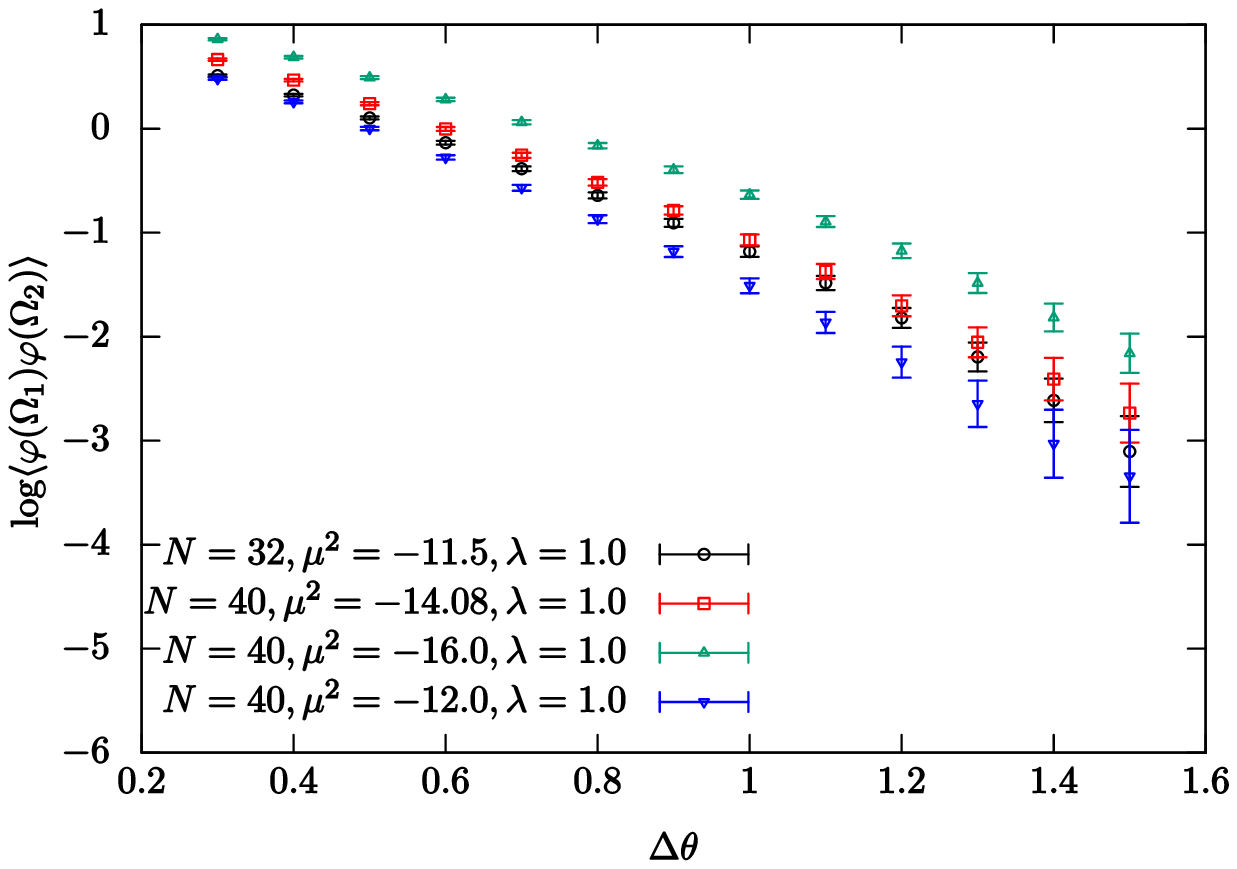}
\caption{$\log\left\langle\varphi(\Omega_1)\varphi(\Omega_2)\right\rangle$
at $\lambda=1.0$ is
plotted against $\Delta\theta$. The data for $N=32$ and $\mu^2=-11.5$ are 
represented by the circles, while the data for
$N=40$ and $\mu^2=-14.08, -16.0, -12.0$ are represented by
the squares, the triangles and the inverted triangles, respectively. }
\label{log_2pt_40}
\end{center}
\end{figure}

\begin{figure}[h]
\begin{center}
\includegraphics[width=11.0cm]{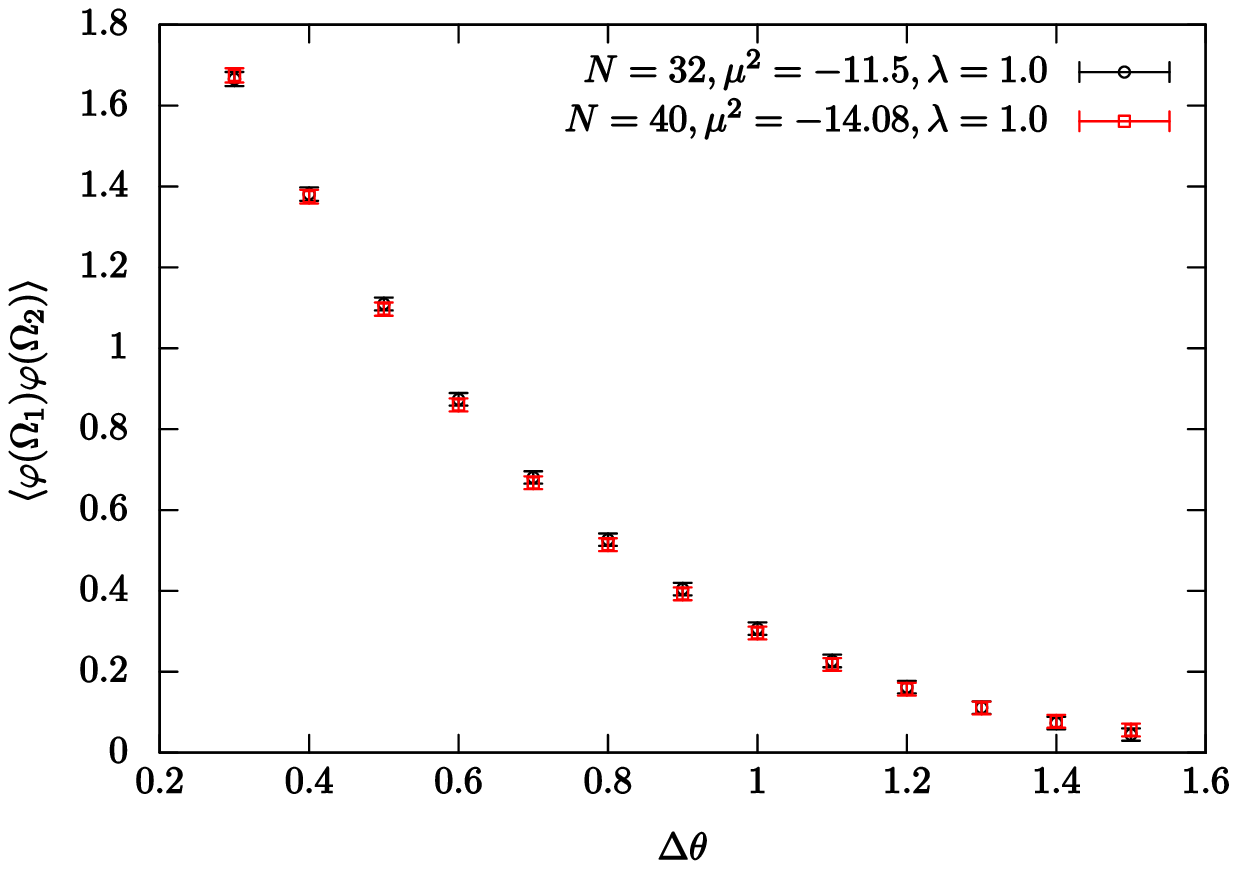}
\caption{$\left\langle\varphi(\Omega_1)\varphi(\Omega_2)\right\rangle$
at $N=32$, $\mu^2=-11.5$ and $\lambda=1.0$ is 
plotted against $\Delta\theta$, where the data are represented by the circles
$\zeta_{40\rightarrow 32}
\left\langle\varphi(\Omega_1)\varphi(\Omega_2)\right\rangle$
with $\zeta_{40\rightarrow 32}=0.8617$
at $N=40$, $\mu^2=-14.08$ and $\lambda=1.0$ 
is also plotted against $\Delta\theta$, where the data are represented by
the squares.}
\label{exp_2pt_40}
\end{center}
\end{figure}

\begin{figure}[h]
\begin{center}
\includegraphics[width=11.0cm]{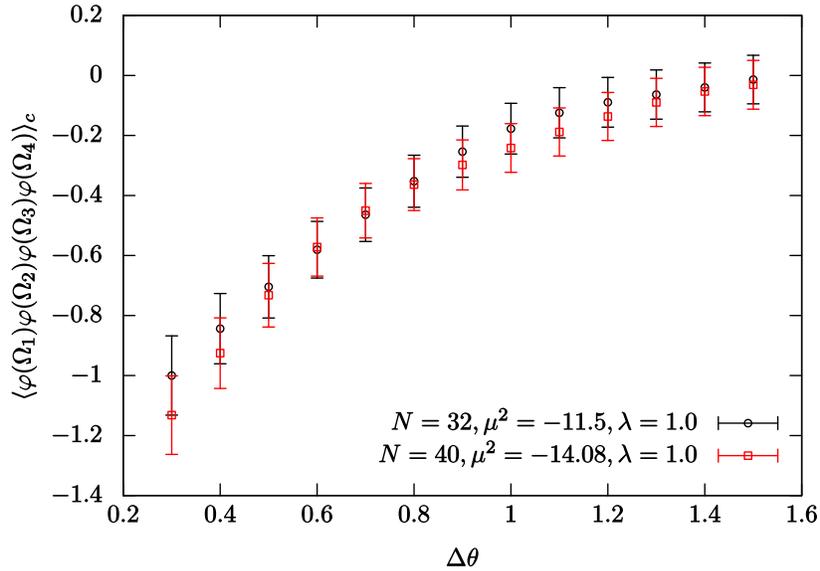}
\caption{$\left\langle \varphi(\Omega_{1}) \varphi(\Omega_{2}) 
\varphi(\Omega_{3}) \varphi(\Omega_{4}) \right\rangle_c$ 
at $N=32$, $\mu^2=-11.5$ and $\lambda=1.0$ is 
plotted against $\Delta\theta$, where the data are represented by the circles.
$\zeta_{40\rightarrow 32}^2
\left\langle \varphi(\Omega_{1}) \varphi(\Omega_{2}) 
\varphi(\Omega_{3}) \varphi(\Omega_{4}) \right\rangle_c$
with $\zeta_{40\rightarrow 32}^2=0.7425$
at $N=40$, $\mu^2=-14.08$ and $\lambda=1.0$ 
is also plotted against $\Delta\theta$, where the data are represented by
the squares.}
\label{exp_4pt_40}
\end{center}
\end{figure}

Similarly, we simulate at $N=40$ and various values of $\mu^2$, 
keeping $\lambda=1.0$.
We perform the same analyses for $N=40$ and $N=32$
in Fig.\ref{log_2pt_40}, Fig.\ref{exp_2pt_40}, and Fig.\ref{exp_4pt_40} as
for $N=24$ and $N=32$
in Fig.\ref{log_2pt_24}, Fig.\ref{exp_2pt_24}, and Fig.\ref{exp_4pt_24}, respectively.
In Fig.\ref{log_2pt_40}, we show the results 
for $\log\left\langle\varphi(\Omega_1)\varphi(\Omega_2)\right\rangle$
at $N=40$ and typical values of $\mu^2$, 
$-14.08, -16.0, -12.0$.
We find that $N=40$ and $\mu^2=-14.08$ corresponds to $N=32$ and 
$\mu^2=-11.5$.
In Fig.\ref{exp_2pt_40} and Fig.\ref{exp_4pt_40}, we confirm that the 2-point function
and the connected 4-point function 
at $N=40$ and $\mu^2=-14.08$ with the wave function renormalization 
agree with those at $N=32$ and $\mu^2=-11.5$.

The values of $\alpha$ and $\zeta$ that we have determined 
are summarized in Table\ref{alpha and zeta}.
The above results strongly suggest that the renormalized correlation functions are
independent of $N$ and that the theory (\ref{action}) is nonperturbatively renormalizable 
in the ordinary sense.

\begin{table}
\centering
\begin{tabular}{ccccc} \toprule
$N' \rightarrow N$ & $\alpha_{N' \rightarrow N}$ & $\delta \alpha_{N' \rightarrow N}$ & $\zeta_{N' \rightarrow N}$ & $\zeta_{N' \rightarrow N}^2$ \\ \midrule
$24 \rightarrow 32$ & $0.2334$ & $0.0108$ & $1.263$ & $1.595$\\
$40 \rightarrow 32$ & $-0.1489$ & $0.0101$ & $0.8617$ & $0.7425$\\ \bottomrule
\end{tabular}
\caption{$\alpha_{N' \rightarrow N}$, $\delta \alpha_{N' \rightarrow N}$, $\zeta_{N' \rightarrow N}$ and $\zeta_{N' \rightarrow N}^2$.}
\label{alpha and zeta}
\end{table}

\newpage

\clearpage

\subsection{One-parameter fine tuning}
In the previous subsection, fixing $\lambda$ and changing the UV cutoff $N$,
we tuned $\mu^2$ depending on $N$ to perform the renormalization. 
The results suggest that tuning a parameter specifies a theory.
Thus we expect that by fixing $N$ and changing $\lambda$ 
one obtains the same theory by tuning $\mu^2$ depending on $\lambda$.

\begin{figure}[h]
\begin{center}
\includegraphics[width=11.0cm]{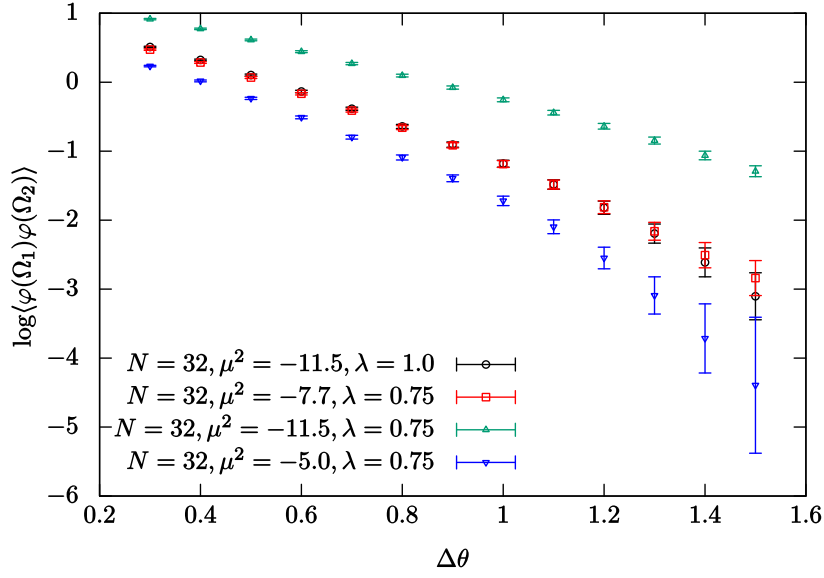}
\caption{$\log\left\langle\varphi(\Omega_1)\varphi(\Omega_2)\right\rangle$
at $N=32$ is
plotted against $\Delta\theta$. The data for $\mu^2=-11.5$ and $\lambda=1.0$ are 
represented by the circles, while the data for
$\mu^2=-7.7, -11.5, -5.0$ and $\lambda=0.75$ are represented by
the squares, the triangles and the inverted triangles, respectively. }
\label{log_2pt_32}
\end{center}
\end{figure}

In this subsection, we see that this is indeed the case.
We simulate at various values of $\mu^2$, $\lambda=0.75$ and $N=32$, and compare
the results with thoes at $\mu^2=-11.5$, $\lambda=1.0$ and $N=32$.
In Fig.\ref{log_2pt_32}, we plot the logarithm of the 2-point functions at 
$\mu^2=-11.5$ and $\lambda=1.0$ and at typical values of $\mu^2$,
$-7.7, -11.5, -5.0$, and $\lambda=0.75$.
We find that the data for $\mu^2=-7.7$ and $\lambda=0.75$ agree with those
for $\mu^2=-11.5$ and $\lambda=1.0$ if the former are simultaneously shifted 
in the vertical direction. By using the least-squares method, we determine that 
$\alpha_{0.75\rightarrow 1.0}=\log(Z(\lambda=1.0)/Z(\lambda=0.75))$.
Correspondingly, we obtain
$\zeta_{0.75\rightarrow 1.0}=e^{\alpha_{0.75\rightarrow 1.0}}
=Z(\lambda=1.0)/Z(\lambda=0.75)$.
The values of $\alpha$ and $\zeta$ that we have determined are summarized
in Table\ref{different lambda}.

\begin{figure}[h]
\begin{center}
\includegraphics[width=11.0cm]{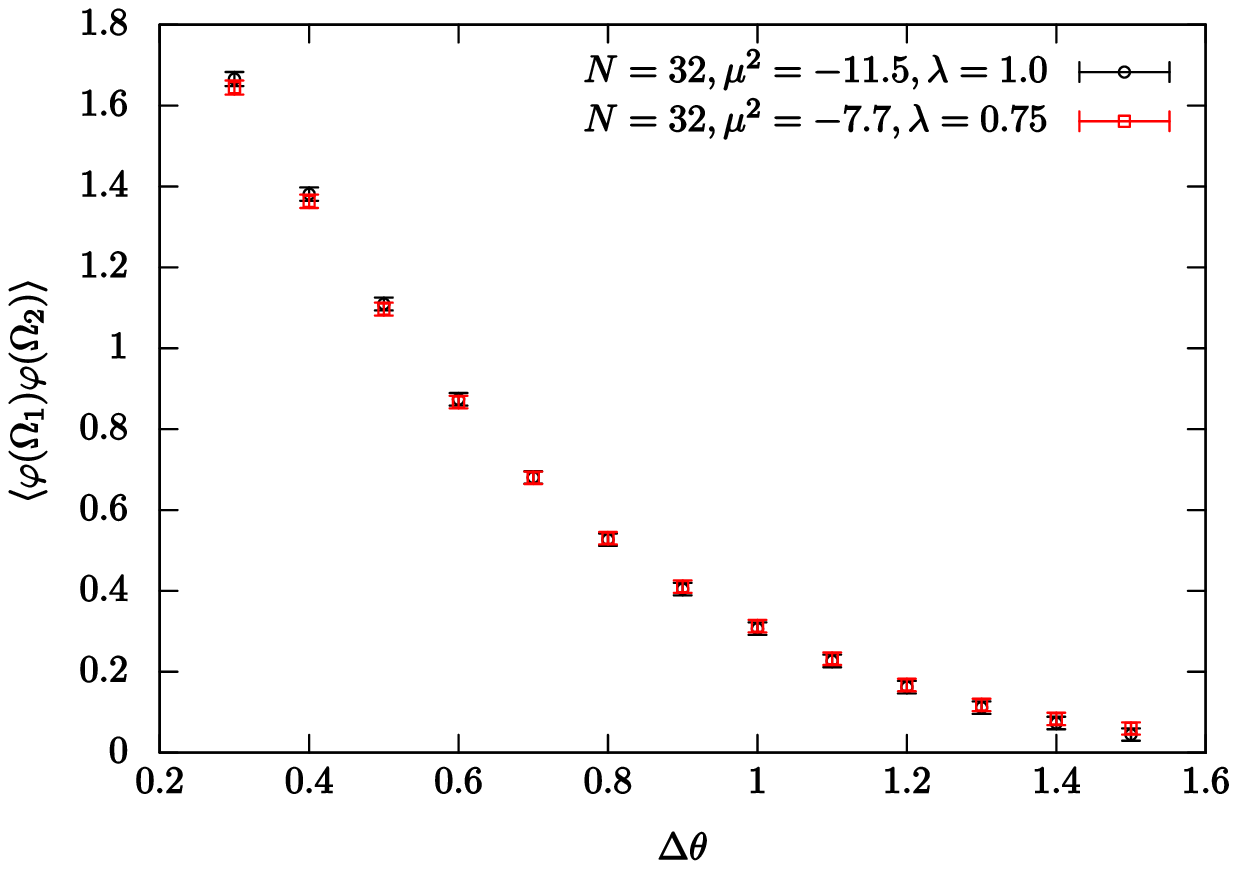}
\caption{$\left\langle\varphi(\Omega_1)\varphi(\Omega_2)\right\rangle$
at $N=32$, $\mu^2=-11.5$ and $\lambda=1.0$ is 
plotted against $\Delta\theta$, where the data are represented by the circles.
$\zeta_{0.75\rightarrow 1.0}
\left\langle\varphi(\Omega_1)\varphi(\Omega_2)\right\rangle$
with $\zeta_{0.75\rightarrow 1.0}=1.025$
at $N=32$, $\mu^2=-7.7$ and $\lambda=0.75$ 
is also plotted against $\Delta\theta$, where the data are represented by
the squares.}
\label{exp_2pt_32}
\end{center}
\end{figure}

\begin{figure}[h]
\begin{center}
\includegraphics[width=11.0cm]{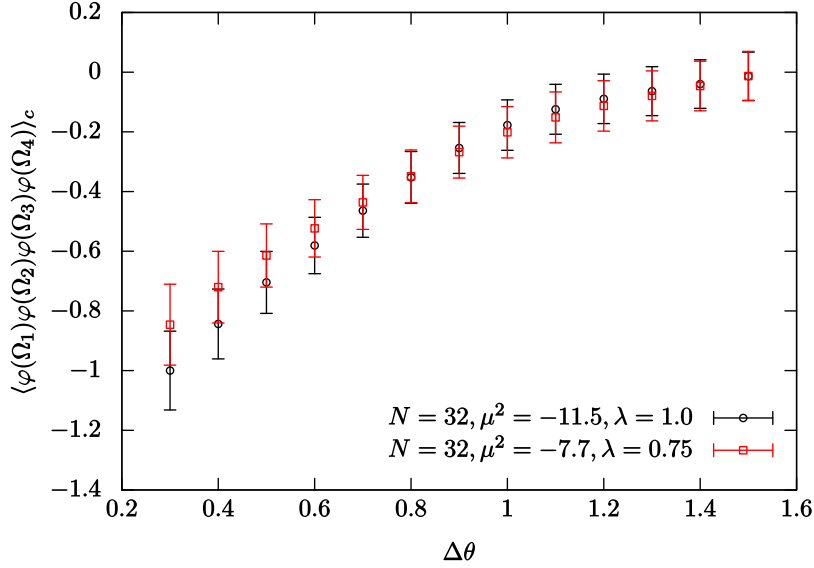}
\caption{$\left\langle \varphi(\Omega_{1}) \varphi(\Omega_{2}) 
\varphi(\Omega_{3}) \varphi(\Omega_{4}) \right\rangle_c$ 
at $N=32$, $\mu^2=-11.5$ and $\lambda=1.0$ is 
plotted against $\Delta\theta$, where the data are represented by the circles.
$\zeta_{0.75\rightarrow 1.0}^2
\left\langle \varphi(\Omega_{1}) \varphi(\Omega_{2}) 
\varphi(\Omega_{3}) \varphi(\Omega_{4}) \right\rangle_c$
with $\zeta_{0.75\rightarrow 1.0}^2=1.051$
at $N=32$, $\mu^2=-7.7$ and $\lambda=0.75$ 
is also plotted against $\Delta\theta$, where the data are represented by
the squares.}
\label{exp_4pt_32}
\end{center}
\end{figure}

In Fig.\ref{exp_2pt_32}, we see that
the 2-point function at $\mu^2=-7.7$ and $\lambda=0.75$ multiplied by 
$\zeta_{0.75\rightarrow 1.0}$ agrees with that at $\mu^2=-11.5$ and $\lambda=1.0$.
In Fig.\ref{exp_4pt_32}, we also see that
the connected 4-point function at $\mu^2=-7.7$ and $\lambda=0.75$ multiplied
by $\zeta_{0.75\rightarrow 1.0}^2$ agrees with that at 
$\mu^2=-11.5$ and $\lambda=1.0$.
These results strongly suggest 
that the theory at $N=32$, $\mu^2=-7.7$ and $\lambda=0.75$
is the same as that at $N=32$, $\mu^2=-11.5$ and $\lambda=1.0$
and that our conjecture that a theory is specified by tuning a parameter is valid.

\begin{table}
\centering
\begin{tabular}{ccccc} \toprule
$\lambda' \rightarrow \lambda$ & $\alpha_{\lambda' \rightarrow \lambda}$ & $\delta \alpha_{\lambda' \rightarrow \lambda}$ & $\zeta_{\lambda' \rightarrow \lambda}$ & $\zeta_{\lambda' \rightarrow \lambda}^2$\\ \midrule
$0.75 \rightarrow 1.0$ & $0.02426$ & $0.01086$ & $1.025$ & $1.051$ \\ \bottomrule
\end{tabular}
\caption{$\alpha_{\lambda' \rightarrow \lambda}$, $\delta \alpha_{\lambda' \rightarrow \lambda}$, $\zeta_{\lambda' \rightarrow \lambda}$ and $\zeta_{\lambda' \rightarrow \lambda}^2$ ($N=32$).}
\label{different lambda}
\end{table}

\clearpage


\section{Conclusion and discussion}
\setcounter{equation}{0}
In this paper, we have studied nonperturbative renormalization 
in a scalar field theory on the fuzzy sphere
by calculating the correlation functions 
by Monte Carlo simulation. The theory is realized by a matrix model, where
the matrix size plays the role of the UV cutoff, and the Berezin symbol constructed 
from the coherent state is identified with the field.
We found that the 2-point and connected 4-point functions 
are made independent of the matrix size
by tuning a parameter and performing the wave function renormalization.
Thus the results strongly suggest that the theory is 
nonperturbatively renormalized  
in the ordinary manner and that the theory is fixed by tuning a parameter.
To support the latter statement,
we examined the correlation functions at fixed $N$.
We found that two different sets of $(\mu^2,\lambda)$ 
indeed give the same theory.

We omitted data for the values of $\Delta\theta$ less than 0.3.
We found that the agreement for $\Delta\theta < 0.3$
between the correlation functions is not so good as that for 
$\Delta\theta \geq 0.3$ while the data for $\Delta\theta < 0.3$ in the 
correlation functions that we compare are very close.
This should be attributed to the effect of the finite UV cutoff, $N$.
We expect the slight deviation for $\Delta\theta < 0.3$ to vanish in $N\rightarrow\infty$.

Now let us see that the $N$ dependence of $\mu^2$ with $\lambda$ fixed is 
approximately given by
\begin{align}
\mu^2=\tilde{\mu}^2-c\log N \ .
\label{equation for mu^2}
\end{align}
By applying $(\mu^2, N)=(-7.97, 24), (-11.5,32)$ to this equation, we obtain
\begin{align}
\tilde{\mu}^2=31.0, \;\;\; c=12.3 \ .
\end{align}
Substituting $N=40$ into (\ref{equation for mu^2}) with the above
values of $\tilde{\mu}^2$
and $c$ yields 
\begin{align}
\mu^2=-14.4 \ ,
\end{align}
which is close to $-14.08$ that we adopted for $N=40$.
Thus there should be a correction to 
(\ref{equation for mu^2}) that vanishes
in the $N\rightarrow\infty$ limit.
On the other hand, 
the one-loop calculation of the self-energy 
in \cite{Chu:2001xi,Steinacker:2016nsc} gives
$c=6\lambda=6.0$.
Hence, our renormalization is indeed a nonperturbative one.
To fix the $N$ dependence of $Z$, we need to further simulate at other $N$'s.

The theory that we have obtained is considered to be 
a finite-volume and noncommutative analog of
the $\lambda\phi^4$ theory in $R^2$, which is obtained by
tuning a parameter and which belongs to the same 
universality class as the 2d Ising model (see, for example, \cite{Loinaz:1997az,
Bosetti:2015lsa}.). 
We should reveal the differences between our theory and an $S^2$ analog of
the $\lambda\phi^4$ theory in $R^2$, whose action is given by
 (\ref{continuum action}), by calculating the correlation functions in the
latter theory and comparing them with thoes that we have obtained in this paper.
Namely, it is important to elucidate 
how noncommutativity or nonlocality affects the correlation functions.

It is shown in
\cite{Martin:2004un,Panero:2006bx,Panero:2006cs,GarciaFlores:2009hf,Das:2007gm} 
that there are three phases in the matrix model (\ref{action}):
the disordered, uniformly ordered and 
striped phases \cite{Gubser:2000cd,Ambjorn:2002nj}.
As we showed in appendix B, the 1-point functions vanish.
This implies that the theory that we have obtained is in the disordered phase.
Indeed, the parameters $\mu^2$ and $\lambda$ that we have used in this paper
are consistent with the disordered phase. 
We would like to examine renormalization in the other phases.
For this purpose, it seems that we need to study a different scaling limit rather
than the commutative limit.
To investigate nonperturbative renormalization in gauge theories
on the fuzzy sphere 
should also be important.

To study the above issues, 
it should be useful to use to other methods
such as the renormalization group analysis developed in \cite{Kawamoto:2015qla}
as well as Monte Carlo simulation.

We hope to report on a study of the above issues in the near future.

\section*{Acknowledgements}
We would like to thank H. Kawai, T. Kuroki, H. Steinacker 
and J. Nishimura for discussions.
Numerical computation was carried out on XC40 at YITP in Kyoto University.
The work of A.T. is supported in part by Grant-in-Aid
for Scientific Research
(No. 15K05046)
from JSPS.


\section*{Appendix A: Bloch coherent state and Berezin symbol}
\setcounter{equation}{0}
\renewcommand{\theequation}{A.\arabic{equation}}
In this appendix, we review the Bloch coherent state \cite{Gazeau},
the Berezin symbol\cite{Berezin:1974du} and the star product on the fuzzy sphere.

We use a standard basis $|jm\rangle$ $(m=-j, -j+1,\dots,j)$
for the spin $j$ representation of the $SU(2)$ algebra,
which obeys the relations
\begin{align}
L_{\pm}|jm\rangle &=\sqrt{(j\mp m)(j \pm m+1)}|j m\pm 1\rangle, \nonumber\\
L_3|jm\rangle &= m |jm\rangle \ ,
\end{align}
where $L_{\pm}=L_1 \pm i L_2$.
The state $|jj\rangle$ is interpreted as corresponding to the north pole on the sphere.
Then, the state $|\Omega\rangle$ 
corresponding to a point $\Omega=(\theta,\varphi)$ is obtained
by multiplying $|jj\rangle$ by a rotation operator as
\begin{align}
|\Omega\rangle=e^{i\theta (\sin\varphi L_1 -\cos\varphi L_2)}|jj\rangle \ .
\label{definition of coherent state}
\end{align}
The state $|\Omega\rangle$ is called the Bloch coherent state. 
It follows from (\ref{definition of coherent state}) that
\begin{align}
n_iL_i |\Omega\rangle =j |\Omega\rangle \ ,
\label{property 1}
\end{align}
where $\vec{n}=(\sin\theta\cos\varphi,\sin\theta\sin\varphi,\cos\theta)$.
(\ref{property 1}) implies that the states $|\Omega\rangle$ 
minimize $\sum_i (\Delta L_i)^2$, where $(\Delta L_i)^2$
is the standard deviation of $L_i$.

Here we introduce the stereographic projection given 
by $z=\tan \frac{\theta}{2} e^{i\varphi}$.
Then, (\ref{definition of coherent state}) is expressed as
\begin{align}
|\Omega\rangle=e^{zL_-} e^{-L_3 \log (1+|z|^2)} e^{-\bar{z}L_+} |jj\rangle \ .
\label{definition of coherent state 2}
\end{align}
By using (\ref{definition of coherent state 2}), an explicit form of
$|\Omega\rangle$ is obtained as
\begin{align}
|\Omega\rangle=\sum_{m=-j}^{j}
\left( 
\begin{array}{c}
2j \\
j+m
\end{array}
\right)^{\frac{1}{2}}
\left( \cos \frac{\theta}{2} \right)^{j+m} \left( \sin \frac{\theta}{2} \right)^{j-m} 
e^{i(j-m) \varphi} |jm\rangle \ .
\label{explicit form}
\end{align}
By using (\ref{explicit form}), it is easy to show the following relations:
\begin{align}
& \langle \Omega_1 | \Omega_2 \rangle
=\left( \cos\frac{\theta_1}{2}\cos\frac{\theta_2}{2}+e^{i(\varphi_2-\varphi_1)}
\sin\frac{\theta_1}{2}\sin\frac{\theta_2}{2} \right)^{2j} \ , 
\label{property2}\\
& |\langle \Omega_1 | \Omega_2 \rangle |= \left(\cos \frac{\chi}{2}\right)^{2j} 
\;\;\mbox{with} \;\;
\chi =\arccos (\vec{n}_1\cdot \vec{n}_2) \ , 
\label{property3}\\
& \frac{2j+1}{4\pi} \int d\Omega \ |\Omega\rangle\langle \Omega | =1 \ .
\label{property4}
\end{align}
Putting $\chi=\frac{2}{\sqrt{j}}$ in the RHS of (\ref{property3}) yields
\begin{align}
\left(\cos\frac{\chi}{2}\right)^{2j} \approx\left(1-\frac{1}{2j}\right)^{2j} \approx e^{-1} 
\end{align}
for large $j$.
This implies that the effective width of the Bloch coherent state is given by
$\frac{R}{\sqrt{j}}=\frac{1}{\sqrt{j}}$.

We also denote the Bloch coherent state $|\Omega\rangle$ by $|z\rangle$.
(\ref{explicit form}) and (\ref{property4}) are rewritten as
\begin{align}
&|z\rangle = \left(\frac{z}{1+|z|^2}\right)^j \sum_{m=-j}^j 
\left(
\begin{array}{c}
2j \\
j+m
\end{array}
\right)^{\frac{1}{2}}
\frac{1}{z^m} 
|jm\rangle \ , \label{explicit form 2}\\
&\frac{2j+1}{4\pi}4\int \frac{d^2z}{(1+|z|^2)^2} \ |z\rangle \langle z| =1 \ ,
\label{property4prm}
\end{align}
respectively.

The Berezin symbol for a $(2j+1)\times (2j+1)$ matrix $A$ is defined by
\begin{align}
f_A(\Omega) & = f_A(z,\bar{z}) \nonumber\\
&=\langle \Omega | A | \Omega \rangle \nonumber\\
&=\langle z | A | z \rangle \ .
\end{align}
The star product for $A$ and $B$ is defined by
\begin{align}
f_A\star f_B(\Omega) = f_A\star f_B (z,\bar{z})
=\langle\Omega | AB |\Omega\rangle =\langle z | AB | z \rangle \ .
\end{align}

Here we consider a quantity
\begin{align}
\frac{\langle w | A | z \rangle}{\langle w | z \rangle} \ .
\end{align}
(\ref{explicit form 2}) implies 
that this quantity is holomorphic with respect to $z$ and anti-holomorphic 
with respect to $w$.
Then, it follows that
\begin{align}
\frac{\langle w | A | z \rangle}{\langle w | z \rangle}
&=e^{-w\frac{\partial}{\partial z}}
\frac{\langle w | A | z+w \rangle}{\langle w | z+w \rangle} \nonumber\\
&=e^{-w\frac{\partial}{\partial z}}e^{z\frac{\partial}{\partial w}}
\frac{\langle w | A | w \rangle}{\langle w | w \rangle} \nonumber\\
&=e^{-w\frac{\partial}{\partial z}}e^{z\frac{\partial}{\partial w}}
\langle w | A | w \rangle \nonumber\\
&=e^{-w\frac{\partial}{\partial z}}e^{z\frac{\partial}{\partial w}}
f_A(w,\bar{w}) \ .
\label{holomorphy}
\end{align}
Similarly,
\begin{align}
\frac{\langle z | A | w \rangle}{\langle z | w \rangle}
=e^{-\bar{w}\frac{\partial}{\partial \bar{z}}}e^{\bar{z}\frac{\partial}{\partial \bar{w}}}
f_A(w,\bar{w}) \ .
\label{holomorphy2}
\end{align}
By using (\ref{property4prm}), (\ref{holomorphy}) and (\ref{holomorphy2}),
one can calculate the star product as
\begin{align}
f_A\star f_B(w,\bar{w})
&=\langle w | AB | w \rangle \nonumber\\
&=\frac{2j+1}{4\pi}4\int \frac{d^2z}{(1+|z|^2)^2}
\frac{\langle w | A | z \rangle}{\langle w | z \rangle}
\frac{\langle z | B | w \rangle}{\langle z | w \rangle}
|\langle w | z \rangle |^2 \nonumber\\
&=\frac{2j+1}{4\pi}4\int \frac{d^2z}{(1+|z|^2)^2}
(e^{-w\frac{\partial}{\partial z}}e^{z\frac{\partial}{\partial w}}
f_A(w,\bar{w}) )
(e^{-\bar{w}\frac{\partial}{\partial \bar{z}}}e^{\bar{z}\frac{\partial}{\partial \bar{w}}}
f_B(w,\bar{w}))
|\langle w | z \rangle |^2 \ .
\label{star product}
\end{align}

\section*{Appendix B: One-point functions}
\setcounter{equation}{0}
\renewcommand{\theequation}{B.\arabic{equation}}

\begin{figure}[t]
\begin{center}
\includegraphics[width=11.0cm]{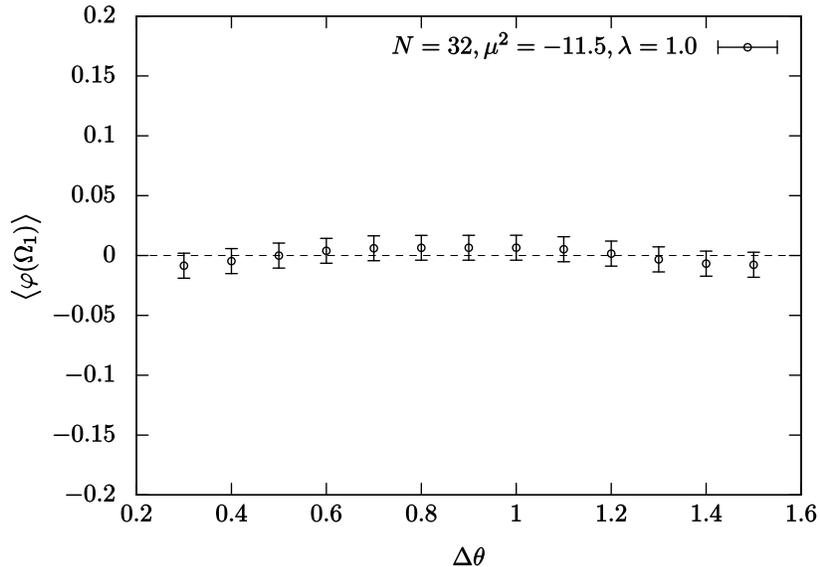}
\caption{$\left\langle\varphi(\Omega_1)\right\rangle$
at $N=32$, $\mu^2=-11.5$ and $\lambda=1.0$ is
plotted against $\Delta\theta$. }
\label{1pt_32}
\end{center}
\end{figure}

In this appendix, we show the results for the 1-point functions.
In Fig.\ref{1pt_32}, we plot $\left\langle\varphi(\Omega_1)\right\rangle$
at $N=32$, $\mu^2=-11.5$ and $\lambda=1.0$
against $\Delta\theta$.
We see that it vanishes within the error.
We have verified that the 1-point functions also vanish
for the other cases that we simulate.

\newpage
\begin{thebibliography}{99}


\bibitem{Douglas:2001ba} 
  M.~R.~Douglas and N.~A.~Nekrasov,
  Rev.\ Mod.\ Phys.\  {\bf 73}, 977 (2001)
  doi:10.1103/RevModPhys.73.977
  [hep-th/0106048].



\bibitem{Connes:1997cr} 
A.~Connes, M.~R.~Douglas and A.~S.~Schwarz,
JHEP {\bf 9802}, 003 (1998)
doi:10.1088/1126-6708/1998/02/003
[hep-th/9711162].


\bibitem{Aoki:1999vr} 
H.~Aoki, N.~Ishibashi, S.~Iso, H.~Kawai, Y.~Kitazawa and T.~Tada,
Nucl.\ Phys.\ B {\bf 565}, 176 (2000)
doi:10.1016/S0550-3213(99)00633-1
[hep-th/9908141].




\bibitem{BFSS}
T.~Banks, W.~Fischler, S.~H.~Shenker and L.~Susskind,
Phys.\ Rev.\ D {\bf 55}, 5112 (1997) [hep-th/9610043].



\bibitem{IKKT}
N.~Ishibashi, H.~Kawai, Y.~Kitazawa and A.~Tsuchiya,
Nucl.\ Phys.\ B {\bf 498}, 467 (1997) [hep-th/9612115].



\bibitem{DVV}
R.~Dijkgraaf, E.~P.~Verlinde and H.~L.~Verlinde,
Nucl.\ Phys.\ B {\bf 500}, 43 (1997) [hep-th/9703030].

\bibitem{Minwalla:1999px} 
S.~Minwalla, M.~Van Raamsdonk and N.~Seiberg,
JHEP {\bf 0002}, 020 (2000)
[hep-th/9912072].
\bibitem{Madore:1991bw} 
J.~Madore,
Class.\ Quant.\ Grav.\ {\bf 9}, 69 (1992).
doi:10.1088/0264-9381/9/1/008

\bibitem{Chu:2001xi} 
C.~S.~Chu, J.~Madore and H.~Steinacker,
JHEP {\bf 0108}, 038 (2001)
[hep-th/0106205].


\bibitem{Steinacker:2016nsc} 
H.~C.~Steinacker,
Nucl.\ Phys.\ B {\bf 910}, 346 (2016)
doi:10.1016/j.nuclphysb.2016.06.029
[arXiv:1606.00646 [hep-th]].

\bibitem{Martin:2004un} 
X.~Martin,
JHEP {\bf 0404}, 077 (2004)
doi:10.1088/1126-6708/2004/04/077
[hep-th/0402230].

\bibitem{Panero:2006bx} 
M.~Panero,
JHEP {\bf 0705}, 082 (2007)
doi:10.1088/1126-6708/2007/05/082
[hep-th/0608202].

\bibitem{Panero:2006cs} 
M.~Panero,
SIGMA {\bf 2}, 081 (2006)
doi:10.3842/SIGMA.2006.081
[hep-th/0609205].

\bibitem{GarciaFlores:2009hf} 
F.~Garcia Flores, X.~Martin and D.~O'Connor,
Int.\ J.\ Mod.\ Phys.\ A {\bf 24}, 3917 (2009)
doi:10.1142/S0217751X09043195
[arXiv:0903.1986 [hep-lat]].

\bibitem{Das:2007gm} 
C.~R.~Das, S.~Digal and T.~R.~Govindarajan,
Mod.\ Phys.\ Lett.\ A {\bf 23}, 1781 (2008)
doi:10.1142/S0217732308025656
[arXiv:0706.0695 [hep-th]].

\bibitem{Kawamoto:2015qla} 
S.~Kawamoto and T.~Kuroki,
JHEP {\bf 1506}, 062 (2015)
[arXiv:1503.08411 [hep-th]].

\bibitem{Vaidya:2003ew} 
S.~Vaidya and B.~Ydri,
Nucl.\ Phys.\ B {\bf 671}, 401 (2003)
doi:10.1016/j.nuclphysb.2003.08.023
[hep-th/0305201].

\bibitem{OConnor:2007ibg} 
D.~O'Connor and C.~Saemann,
JHEP {\bf 0708}, 066 (2007)
doi:10.1088/1126-6708/2007/08/066
[arXiv:0706.2493 [hep-th]].

\bibitem{Nair:2011ux} 
V.~P.~Nair, A.~P.~Polychronakos and J.~Tekel,
Phys.\ Rev.\ D {\bf 85}, 045021 (2012)
doi:10.1103/PhysRevD.85.045021
[arXiv:1109.3349 [hep-th]].

\bibitem{Polychronakos:2013nca} 
A.~P.~Polychronakos,
Phys.\ Rev.\ D {\bf 88}, 065010 (2013)
doi:10.1103/PhysRevD.88.065010
[arXiv:1306.6645 [hep-th]].

\bibitem{Tekel:2013vz} 
J.~Tekel,
Phys.\ Rev.\ D {\bf 87}, no. 8, 085015 (2013)
doi:10.1103/PhysRevD.87.085015
[arXiv:1301.2154 [hep-th]].

\bibitem{Saemann:2014pca} 
C.~Saemann,
JHEP {\bf 1504}, 044 (2015)
doi:10.1007/JHEP04(2015)044
[arXiv:1412.6255 [hep-th]].

\bibitem{Tekel:2014bta} 
J.~Tekel,
JHEP {\bf 1410}, 144 (2014)
doi:10.1007/JHEP10(2014)144
[arXiv:1407.4061 [hep-th]].

\bibitem{Tekel:2015zga} 
J.~Tekel,
JHEP {\bf 1512}, 176 (2015)
doi:10.1007/JHEP12(2015)176
[arXiv:1510.07496 [hep-th]].

\bibitem{Bietenholz:2004xs} 
W.~Bietenholz, F.~Hofheinz and J.~Nishimura,
JHEP {\bf 0406}, 042 (2004)
doi:10.1088/1126-6708/2004/06/042
[hep-th/0404020].
\bibitem{Mejia-Diaz:2014lza} 
H.~Mej\'{i}a-D\'{i}az, W.~Bietenholz and M.~Panero,
JHEP {\bf 1410}, 56 (2014)
doi:10.1007/JHEP10(2014)056
[arXiv:1403.3318 [hep-lat]].

\bibitem{Bietenholz:2002ch} 
W.~Bietenholz, F.~Hofheinz and J.~Nishimura,
JHEP {\bf 0209}, 009 (2002)
doi:10.1088/1126-6708/2002/09/009
[hep-th/0203151].

\bibitem{Bietenholz:2006cz} 
W.~Bietenholz, J.~Nishimura, Y.~Susaki and J.~Volkholz,
JHEP {\bf 0610}, 042 (2006)
doi:10.1088/1126-6708/2006/10/042
[hep-th/0608072].

\bibitem{Berezin:1974du} 
F.~A.~Berezin,
Commun.\ Math.\ Phys.\ {\bf 40}, 153 (1975).

\bibitem{Gazeau}
J. P. Gazeau. Coherent states in quantum physics - 2009. Weinheim, Germany: WileyVCH. 

\bibitem{Iso:2000ew} 
S.~Iso, H.~Kawai and Y.~Kitazawa,
Nucl.\ Phys.\ B {\bf 576}, 375 (2000)
doi:10.1016/S0550-3213(00)00092-4
[hep-th/0001027].


\bibitem{Okuno:2015kuc} 
S.~Okuno, M.~Suzuki and A.~Tsuchiya,
PTEP {\bf 2016}, no. 2, 023B03 (2016)
doi:10.1093/ptep/ptv192
[arXiv:1512.06484 [hep-th]].

\bibitem{Suzuki:2016sca} 
M.~Suzuki and A.~Tsuchiya,
PTEP {\bf 2017}, no. 4, 043B07 (2017)
[arXiv:1611.06336 [hep-th]].

\bibitem{Karczmarek:2013jca} 
J.~L.~Karczmarek and P.~Sabella-Garnier,
JHEP {\bf 1403}, 129 (2014)
[arXiv:1310.8345 [hep-th]].

\bibitem{Sabella-Garnier:2014fda} 
P.~Sabella-Garnier,
JHEP {\bf 1502}, 063 (2015)
[arXiv:1409.7069 [hep-th]].

%


%




\bibitem{Alexanian:2000uz} 
G.~Alexanian, A.~Pinzul and A.~Stern,
Nucl.\ Phys.\ B {\bf 600}, 531 (2001)
[hep-th/0010187].


\bibitem{Hammou:2001cc} 
A.~B.~Hammou, M.~Lagraa and M.~M.~Sheikh-Jabbari,
Phys.\ Rev.\ D {\bf 66}, 025025 (2002)
[hep-th/0110291].


\bibitem{Presnajder:1999ky} 
P.~Presnajder,
J.\ Math.\ Phys.\ {\bf 41}, 2789 (2000)
[hep-th/9912050].

\bibitem{Ishiki:2015saa} 
G.~Ishiki,
Phys.\ Rev.\ D {\bf 92}, no. 4, 046009 (2015)
[arXiv:1503.01230 [hep-th]].

\bibitem{Loinaz:1997az} 
  W.~Loinaz and R.~S.~Willey,
  Phys.\ Rev.\ D {\bf 58}, 076003 (1998)
  doi:10.1103/PhysRevD.58.076003
  [hep-lat/9712008].

\bibitem{Bosetti:2015lsa} 
  P.~Bosetti, B.~De Palma and M.~Guagnelli,
  Phys.\ Rev.\ D {\bf 92}, no. 3, 034509 (2015)
  doi:10.1103/PhysRevD.92.034509
  [arXiv:1506.08587 [hep-lat]].



\bibitem{Gubser:2000cd} 
S.~S.~Gubser and S.~L.~Sondhi,
Nucl.\ Phys.\ B {\bf 605}, 395 (2001)
doi:10.1016/S0550-3213(01)00108-0
[hep-th/0006119].


\bibitem{Ambjorn:2002nj} 
J.~Ambjorn and S.~Catterall,
Phys.\ Lett.\ B {\bf 549}, 253 (2002)
doi:10.1016/S0370-2693(02)02906-4
[hep-lat/0209106].




\end{thebibliography}
\end{document}